\documentclass[a4paper]{article}
\usepackage[utf8]{inputenc}
\usepackage{graphicx}
\usepackage{onecolceurws}
\usepackage{hyperref}
\usepackage{enumitem}
\setlist{nosep}

\title{Expert-sourcing Domain-specific Knowledge: The Case of Synonym Validation}

\author{
Michael Unterkalmsteiner \\ Software Engineering Research Lab Sweden\\
                Blekinge Institute of Technology \\ 
                Karlskrona, Sweden \\
                michael.unterkalmsteiner@bth.se
\and
Andrew Yates \\ Max Planck Institute for Informatics\\
                Saarbrücken, Germany \\ ayates@mpi-inf.mpg.de
}

\institution{}

\begin{document}
\maketitle

\begin{abstract}
	One prerequisite for supervised machine learning is high quality labelled data. Acquiring such data is, particularly if expert knowledge is required, costly or even impossible if the task needs to be performed by a single expert. In this paper, we illustrate tool support that we adopted and extended to source domain-specific knowledge from experts. We provide insight in design decisions that aim at motivating experts to dedicate their time at performing the labelling task. We are currently using the approach to identify true synonyms from a list of candidate synonyms. The identification of synonyms is important in scenarios were stakeholders from different companies and background need to collaborate, for example when defining and negotiating requirements. We foresee that the approach of expert-sourcing is applicable to any data labelling task in software engineering. The discussed design decisions and implementation are an initial draft that can be extended, refined and validated with further application.
\end{abstract}
\vskip 32pt

\section{Introduction}
The training and validation of natural language processing models, that are based on supervised machine learning, require data that is labelled by humans. Creating labelled data, in particular if it is domain specific, is costly and can require expert knowledge. Furthermore, the lack of high-quality labelled data may prevent the transfer of an approach from one domain to the other, simply because not enough labelled data exists to train the model~\cite{ferrari_natural_2018}. 

Crowdsourcing platforms provide the possibility to harvest human intelligence that can be used for data labelling. While this works well for tasks that target the humans' predisposition for pattern recognition, tasks for which domain-specific knowledge is required cannot be outsourced to an arbitrary crowd. Such tasks need to be designed such that a limited target group remains engaged with the data labelling task and experiences benefits from participation. In this paper, we provide some insight in an ongoing study and provide motivation for the design decisions we made when adopting an existing crowdsourcing tool for our particular task: validation of domain-specific synonym candidates.

\section{Background}
Our current research focuses at supporting requirements engineers to adopt an object classification system, CoClass\footnote{\url{https://coclass.byggtjanst.se/en/about\#about-coclass}}, from the construction business domain. The classification is planned to be used throughout the organization to identify and trace specified, designed, constructed and eventually maintained objects. CoClass is a hierarchical ontology of construction objects that provides a coding system, a definition and synonyms for each object. CoClass is still under development and many object to synonym mappings are still incomplete. These mappings are however important for the use of the classification system as it allows users, with different background and vocabulary, to find the objects they are looking for. Furthermore, we plan to use the ontology to automatically classify natural language requirements such that they can be traced during the life-cycle of a project. 

\subsection{Domain-specific synonym detection}
In order to fill the synonym gaps in CoClass, we use a learning-to-rank approach for domain-specific synonym detection~\cite{yates_replicating_2019}. The basic idea of this supervised approach is to learn term associations from a domain specific corpus, using features that indicate the synonymous use of a term. The approach produces a list of synonym candidates for each term defined in CoClass (1430 terms with each 1000 synonym candidates). A preliminary evaluation of the candidates with a domain expert suggests that only $\sim$1\% of the synonym candidates are true synonyms (10 in 1000). While this precision might seem underwhelming, automated synonym detection \emph{is} difficult and should be compared against its manual alternatives or evaluated against the cost of not discovering new synonyms at all.

\section{Expert-sourcing synonym validation}
Reviewing 1,430,000 synonym candidates would be a monumental task for an individual. While crowdsourcing~\cite{doan_crowdsourcing_2011} the task to the general public would be possible, it would likely not succeed, as the task is language (Swedish) and domain (construction business) specific, limiting the potential and reliable participants considerably. We chose therefore to use a crowdsourcing framework, Pybossa\footnote{\url{https://github.com/Scifabric/pybossa}}, that allows us to control all aspects of the validation process: participants, data storage and task design. Pybossa provides important infrastructure for realizing a crowdsourcing project, such as task importing, management, scheduling, and redundancy, user management and results analysis. In addition, Pybossa provides a REST API and convenience functions that can be used for tasks, e.g. a media player for video/sound annotation tasks or a PDF reader for transcription tasks.
In the remainder of the paper, we focus on the task design and the decisions that were made in order to make the validation process efficient and effective. The code for task presentation and analysis is available online\footnote{\url{https://github.com/munterkalmsteiner/pybossa-trafikverket-theme}}. 

\subsection{Task design}
\begin{figure}[!t]
	\centering
	\includegraphics[width=0.8\textwidth]{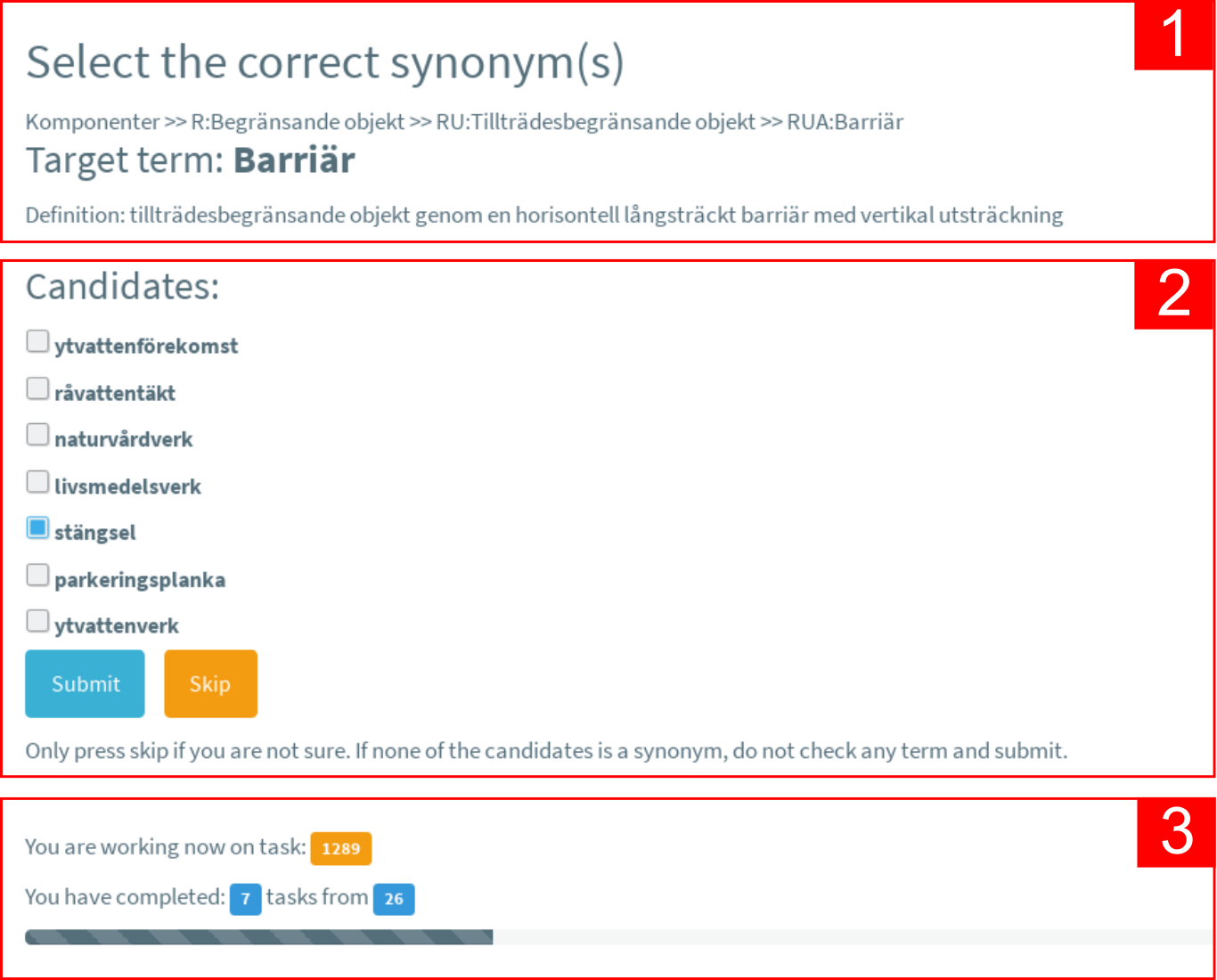}
	\caption{The selection phase of the task}\label{fig:s}
\end{figure}
\begin{figure}[!t]
	\centering
	\includegraphics[width=0.9\textwidth]{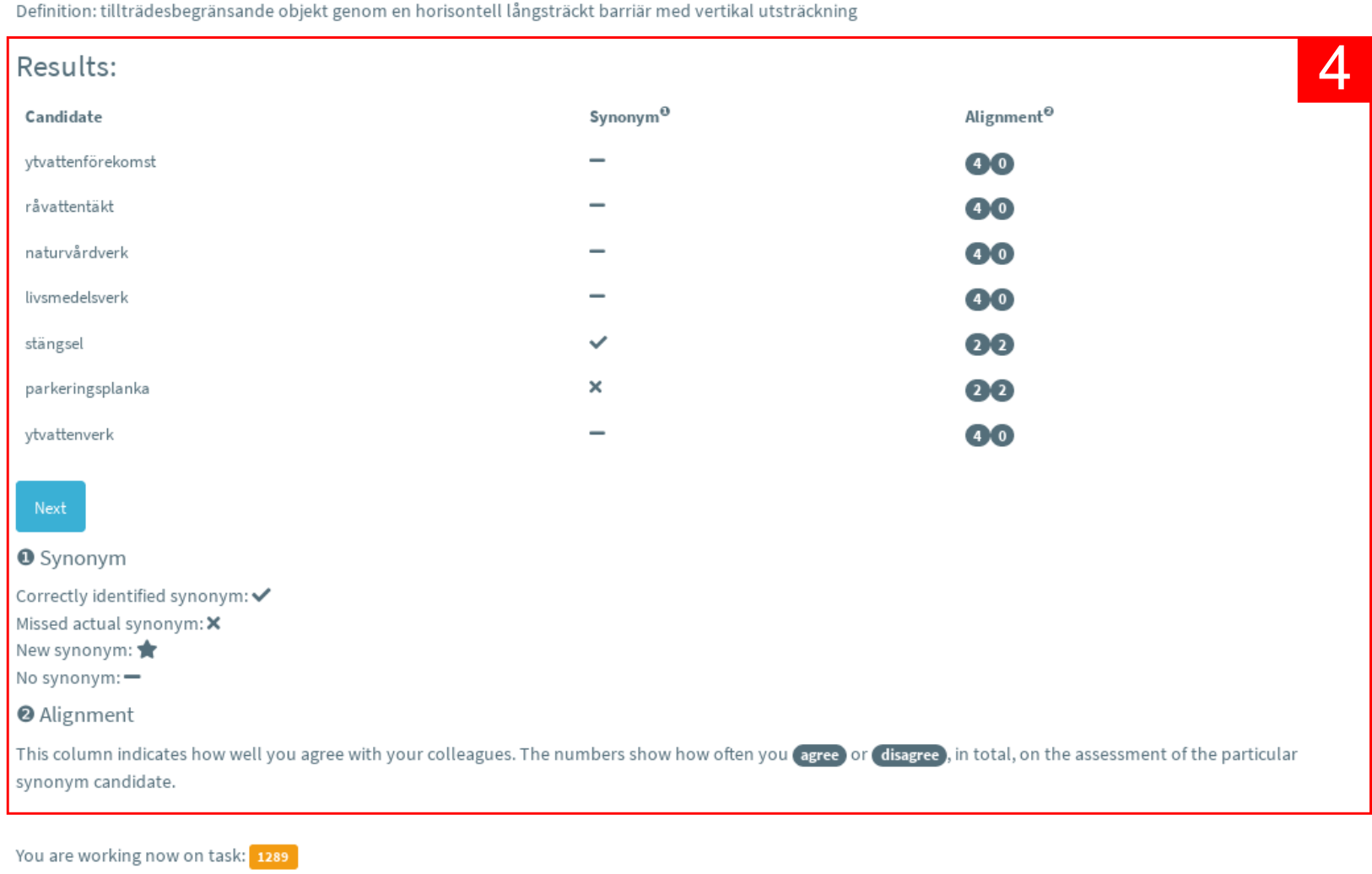}
	\caption{The results phase of the task}\label{fig:r}
\end{figure}
The validation task is separated into two phases. In phase 1, the selection, the expert selects $0..n$ synonyms from a list of candidates for a particular target term. In phase 2, the result, the expert receives feedback on his/her selection. Screenshots of the respective phases are shown in Figure~\ref{fig:s} and~\ref{fig:r}. The red markers are inserted for referencing purposes, used in the following discussion. 

Panel 1 in Figure~\ref{fig:s} shows the target term for which the expert needs to select synonyms. In this area, we show also the hierarchical structure of CoClass under which the target term (transl.: fence) can be found (transl.: Components $\gg$ Limiting objects $\gg$ Access-limiting objects $\gg$ Fence), including the coding that is used for such objects (R $\gg$ RU $\gg$ RUA). We also show the definition used in CoClass of the target term (transl: access restricting object by a horizontal elongated barrier with a vertical extent). The purpose is to provide context, to foster organizational learning~\cite{kim_link_1993} and to develop a common vocabulary that potentially reduces misunderstandings in the organization.

Panel 2 in Figure~\ref{fig:s} shows the list of candidate synonyms. We group candidate synonyms with affinity propagation clustering~\cite{frey_clustering_2007}, measuring similarity with the Levenshtein distance. This reduces the perceived number of terms an expert has to inspect as similar terms can be accepted/rejected in one task. If the expert is not sure about the meaning of the term or the synonym candidates, (s)he can skip the task and proceed to the next one.
Panel 3 in Figure~\ref{fig:s} shows the overall progress, i.e. tasks done of the total number of tasks. 

Once the expert has made a decision, the results for the particular task are stored and analysed in order to provide immediate feedback to the expert. An example of the analysis is shown in Panel 4 in Figure~\ref{fig:r}. In the second column of the results table, we show whether the selected term is a correctly identified or a missed actual synonym, according to the already defined synonyms in CoClass, or a completely new identified synonym. In the third column, we show how well aligned the current expert is with other experts that have already performed the same task. For example, the selection of the expert in Figure~\ref{fig:r} has missed the actual synonym ``parkeringsplanka'', and so did another user. They agree that ``parkeringsplanka'' is not a synonym of ``barriär''. However, two other experts had a different opinion, i.e. ``parkeringsplanka'' is indeed a synonym of ``barriär''.
Once the tasks are completed, it is straightforward to identify new synonyms with a simple majority vote.    

\subsection{Motivational aspects of expert-sourcing}
\begin{figure}
	\centering
	\includegraphics[width=\textwidth]{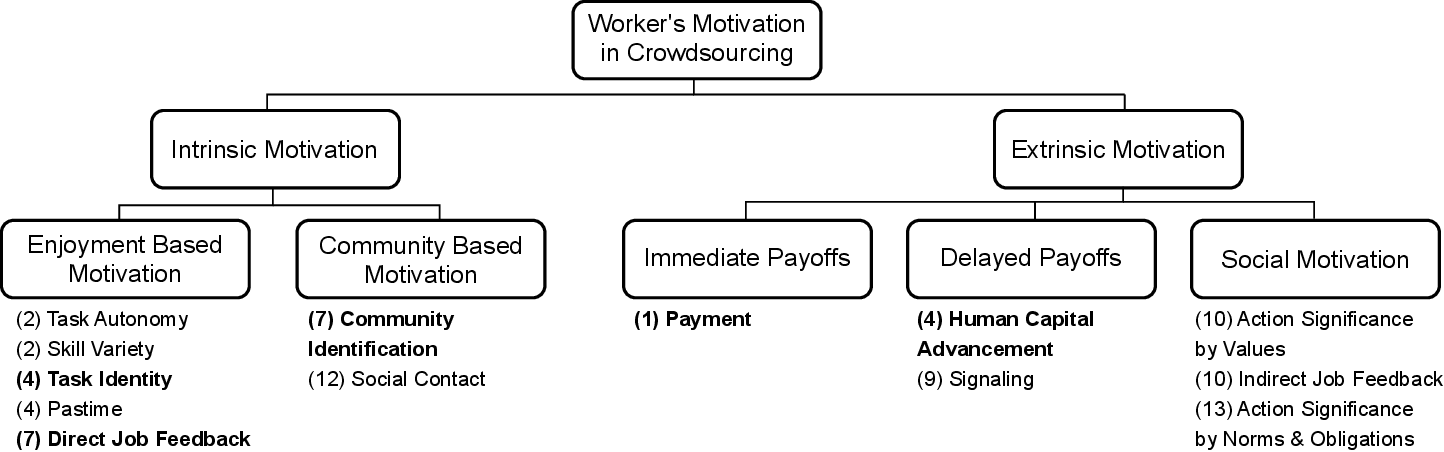}
	\caption{Model for worker's motivation in crowdsourcing, adapted from Kaufmann et al.~\cite{kaufmann_more_2011}}\label{fig:motivation}
\end{figure}
When we designed the task, we considered how to create a win-win situation for the participating experts, management which pays for the time spent on the task, and researchers. There exists some evidence that intrinsic motivation is more important than extrinsic motivation for crowd-sourcing workers~\cite{kaufmann_more_2011}. Figure~\ref{fig:motivation} shows different aspects of motivation and their relative importance ranking (1-13), based on a survey of 431 Amazon Mechanical Turk workers. In the remainder of this section, we discuss our strategies to foster some aspects of worker motivation. 

The synonym selection task should transfer some knowledge to the participants. We provide that by showing term definitions, the CoClass hierarchy and code under which the term is found. This fosters individual learning as well as organizational learning as it promotes a common vocabulary (Human Capital Advancement, i.e. motivation to enable training of skills). 
Similarly, the feedback on the results page helps individuals to understand how well they are aligned with their colleagues (Direct Job Feedback, i.e. motivation provided by the perception of achievement; Community Identification, i.e. the subconscious adoption of norms and values). For management, this could also be useful information as it could indicate where adjustments in documentation or training are needed. Since we know exactly how much time each expert has spent on their tasks, we can quantify the cost for collecting synonyms and potential terminology misalignments (Payment, i.e. motivation by monetary compensation). Such figures can help to get management buy-in when extending the study or replicating it in another organization.

A potential threat to the validation of the synonyms is the result page where we show the alignment of experts immediately after their choice (Task Identity, i.e. the extent to which a participant perceives that his/her work leads to a result). Therefore, we randomize the presentation of tasks (in blocks of five, i.e. after five tasks we change the target term), counteracting conscious or unconscious bias. Finally, we seed a true synonym if the expert did not select a synonym after 10 tasks in a row. The intention is both to keep the participant motivated by ``finding'' a synonym and to verify that the expert is still paying attention to the task and not submitting random answers.

In Figure~\ref{fig:motivation}, we highlight in \textbf{bold} typeface which motivational aspects we address. We briefly discuss which aspects are not covered. Task Autonomy refers to the degree to which creativity and own decisions are permitted by the task. The nature of data labelling tasks leaves little leeway and creativity would rather be counter productive. It would be difficult to design a task that caters for this motivational aspect. Skill Variety refers to the usage of different skills for solving a task that match to the available skill set of the worker. One way to address this motivational aspect would be to segment the CoClass terms into themes that require specialized subdomain knowledge, matching a subset of participants' specialized background and expertise. Pastime refers to the motivation to do something in order to avoid boredom. One could argue that, since the synonym selection task can be performed on mobile devices (e.g. while riding the train to work), this motivational aspect is covered. On the other hand, the task \emph{is} work and part of the professional activities of an employee, making this motivational aspect not applicable to our context. We do not address any aspects from the range of social motivations. Indirect Job Feedback, i.e. motivation through feedback about the delivered work, for example through comments and other encouragements could however be implemented. Finally, we do not use yet any form of gamification mechanisms. Leaderboards and level systems can be effective means to increase long-term engagement and quality of output~\cite{morschheuser_gamification_2016}.  

\section{Conclusions and Future Work}
In this paper, we suggest expert-sourcing as a mean to acquire labelled data from domain experts. We illustrate the adoption of a crowd-sourcing platform and the design of the data labelling task for domain-specific synonym identification such that it is engaging and useful for the participating experts. We are currently in the process of piloting the approach with select domain experts and gather feedback on the task design. Once the task design is stabilized, we intend to deploy the data collection mechanism to approximately 500 participants. 

While we apply the approach to a narrow, specialised, problem (synonym identification), the idea and design decisions to cater for motivational aspects are generally applicable to any data labelling task in Software Engineering. One could design tasks to evaluate the quality of certain artefacts and use this assessment to train a classification algorithm, for example to evaluate the degree of ambiguity in statements of requirements specifications, the understandability of test cases, identification of code refactorings, detection of code smells or the readability of source code. 

\bibliographystyle{alpha} 
\bibliography{references}

\end{document}